# Perfect Nonradiating Modes in Dielectric Nanoparticles

Vasily V. Klimov

P.N. Lebedev Physical Institute, Russian Academy of Sciences,53 Leninsky Prospekt,

Moscow 119991, Russia

E_mail: klimov256@gmail.com

Abstract

A general concept of perfect nonradiating modes in dielectric nanoparticles of an arbitrary shape is put forward. It is proved that such modes exist in axisymmetric dielectric nanoparticles and have unlimited radiative $Q$-factors. With smart tuning of the excitation beams, perfect modes appear as deep minima in the scattered radiation spectra (up to complete disappearance), but at the same time they have a giant amplification of the fields inside the particle. Such modes have no analogues and can be useful for the realization of nanosensors, low threshold nanolasers and other strong nonlinear effects in nanoparticles.

At present, the properties of dielectric nanoparticles with a high refractive index and low radiative losses are being actively investigated [1-9]. The physics of optical phenomena in such nanoparticles is very complicated and leads to many interesting applications, such as nano-antennas [1-3], nanolasers [4,5] and nonlinear nanophotonics [8,9]. As in any other field of physics, all these phenomena are associated with the existence of certain eigenmodes in nanoparticles.

For applications, modes with strong field localization and low radiation losses are of particular interest. This kind of modes had attracted the closest attention of the leading scientific groups, discovered several types of weakly radiating phenomena: BIC modes [10-15], anapole states [8,10,16-23], supercavity modes [5, 24, 25], pseudo-modes [26], embedded photonic eigenvalues [27]. In this letter, we show that there are unparalleled perfect nonradiating modes in dielectric nanoparticles and we propose a regular method for finding such modes in arbitrary dielectric particles.

Usually, the eigenmodes are found by solving sourceless Maxwell's equations with the Sommerfeld radiation conditions at infinity [28], and therefore such modes are fundamentally related to radiation losses. Moreover, such modes grow unlimitedly at infinity, requiring the development of very complex artificial approaches for their use, see e.g. [29-36].

However, finding all the modes is a non-trivial task, not only from a computational point of view, and the modes investigated in the abovementioned works do not exhaust the entire set of modes that exist in dielectric particles. In this paper, we present a new class of eigenmodes - perfect nonradiating modes, and, to find them, we propose to use the solutions of Maxwell's equations, not containing waves that carry energy away in principle!

More specifically, we suggest to look for the electromagnetic fields outside the particle in the form of a superposition of solutions of the Maxwell's equations that are nonsingular in unlimited free space, including the interior of the nanoparticle. This approach is fundamentally different from the usual one, assuming that the functions describing the fields outside the body can have singularities upon analytic continuation into its interior. For example, expanding any field component $E(r,\theta,\varphi,\omega)$ outside the resonator over spherical harmonics $Y_n^m(\theta,\varphi)$ in accordance with the Sommerfeld radiation condition, usually one uses spherical Hankel functions $h_n^{(1)}(k_0 r)$ singular at $r = 0$ (inside the resonator):

$$E(r,\theta,\varphi,\omega) \sim \sum a_{nm} h_n^{(1)}(k_0 r) Y_n^m(\theta,\varphi) \quad (1),$$

fundamentally related to radiation. We propose, however, to find the perfect nonradiating modes to use only the components nonsingular at the origin to describe the fields at infinity

$$E(r,\theta,\varphi,\omega) \sim \sum a_{nm} j_n(k_0 r) Y_n^m(\theta,\varphi) \quad (2)$$

where $j_n(k_0 r)$ are nonsingular spherical Bessel functions. The choice of the asymptotics (2) is directly associated with the peculiarities of the expansion of any incident beams over the spherical harmonics $Y_n^m(\theta,\varphi)$ near the scattering bodies: such an expansion does not contain the Neumann functions singular at the origin located inside the resonator. For example, the expansion of a plane wave near a scattering body has the form [37]:

$$e^{ik_0 R\cos\theta} = \sum_{n=0}^{\infty} i^n (2n+1) j_n(k_0 R) P_n(\cos\theta) \quad (3),$$

that does not contain singular at the origin Neumann functions $y_n(k_0 R)$. In (3), $P_n(\cos\theta)$ are Legendre polynomials

Obviously, if a solution of the Maxwell's equations with asymptotics (2) exists, then, in principle, it does not have a flux of energy and radiation. Since (2) has no singularities in the entire space (including the nanoparticle interior), finding of modes in infinite space can be reduced to finding fields in the volume of a nanoparticle only. As a result, the system of equations that determine unambiguously the perfect nonradiating modes in a nonmagnetic nanoparticle with the permittivity $\varepsilon$ can be written as a system of two equations for two fields $\mathbf{E}_1$, $\mathbf{E}_2$ inside the particle:

$$\begin{aligned}\nabla \times \mathbf{E}_1 = ik_0 \mathbf{H}_1; \nabla \times \mathbf{H}_1 = -ik_0 \varepsilon \mathbf{E}_1, \text{inside nanoparticle} \\ \nabla \times \mathbf{E}_2 = ik_0 \mathbf{H}_2; \nabla \times \mathbf{H}_2 = -ik_0 \mathbf{E}_2, \text{inside nanoparticle}\end{aligned} \quad (4)$$

connected through the conditions of continuity of the tangential components of the electric and magnetic fields at the nanoparticle boundary:

$$\mathbf{E}_{t,1} = \mathbf{E}_{t,2}; \mathbf{H}_{t,1} = \mathbf{H}_{t,2} \quad (5)$$

It is important to note that the system (4) and (5) is self-sufficient, and there is no need to impose any condition at infinity to solve it.

At some real values of frequency $k_{0,n}$, the system of equations (4), (5) becomes compatible, that is, the perfect nonradiating modes appear. It is very important that due to a specific structure of (4) and (5) there is nothing common between the frequencies of perfect modes and the frequencies of usual normal modes. The modes found from (4), (5) are orthogonal in the sense that

$$\begin{aligned}\int_V dV \left( \mathbf{H}_{1,n} \cdot \mathbf{H}_{1,m} - \mathbf{H}_{2,n} \cdot \mathbf{H}_{2,m} \right) = \delta_{nm} \\ \int_V dV \left( \varepsilon \mathbf{E}_{1,n} \cdot \mathbf{E}_{1,m} - \mathbf{E}_{2,n} \cdot \mathbf{E}_{2,m} \right) = \delta_{nm}\end{aligned} \quad (6),$$

where the integration is over the resonator volume $V$. The condition (6) is also drastically different from orthogonality conditions for usual normal modes.

The fields inside the nanoparticle are determined by $\mathbf{E}_1, \mathbf{H}_1$, while the fields outside the particle are determined by the analytic continuation of the solution $\mathbf{E}_2, \mathbf{H}_2$.

The system of equations (4),(5) is very complicated, and a rigorous mathematical theory in a general case does not yet exist for it. Nevertheless, we managed to find conditions for the existence of perfect nonradiating modes for arbitrary spheres, spheroids and super-spheroids, describing well practically all forms of nanoparticles interesting for applications.

First of all, perfect nonradiating modes exist for a spherical particle of the radius $R$, having the following solution of the system (4),(5) for TM polarization:

$$H_\varphi = j_n(z_0) j_n(k_0\sqrt{\varepsilon}r) P_n^1(\cos\theta), r < R$$
$$H_\varphi = j_n(z_1) j_n(k_0 r) P_n^1(\cos\theta), r > R \qquad (7)$$
$$z_0 = k_0 R; z_1 = k_0\sqrt{\varepsilon}R$$

where $P_n^1(\cos\theta)$ are Legendre polynomials.

Note that the condition for the existence of perfect nonradiating modes (7) has the form

$$\varepsilon j_n(z_1)[z_0 j_n(z_0)]' = j_n^{(1)}(z_0)[z_1 j_n(z_1)]' \qquad (8)$$

and coincides with the vanishing of the numerator of the Mie scattering coefficient. Dispersion equation (8), along with complex roots, also has real roots, corresponding to perfect nonradiating modes. Fig. 1 shows the dependences of $\text{Re}(H_\varphi(r, \theta = \pi/2))$ on the radius for a perfect nonradiating mode and a usual mode with radiation losses in a sphere with $\varepsilon = 10$.

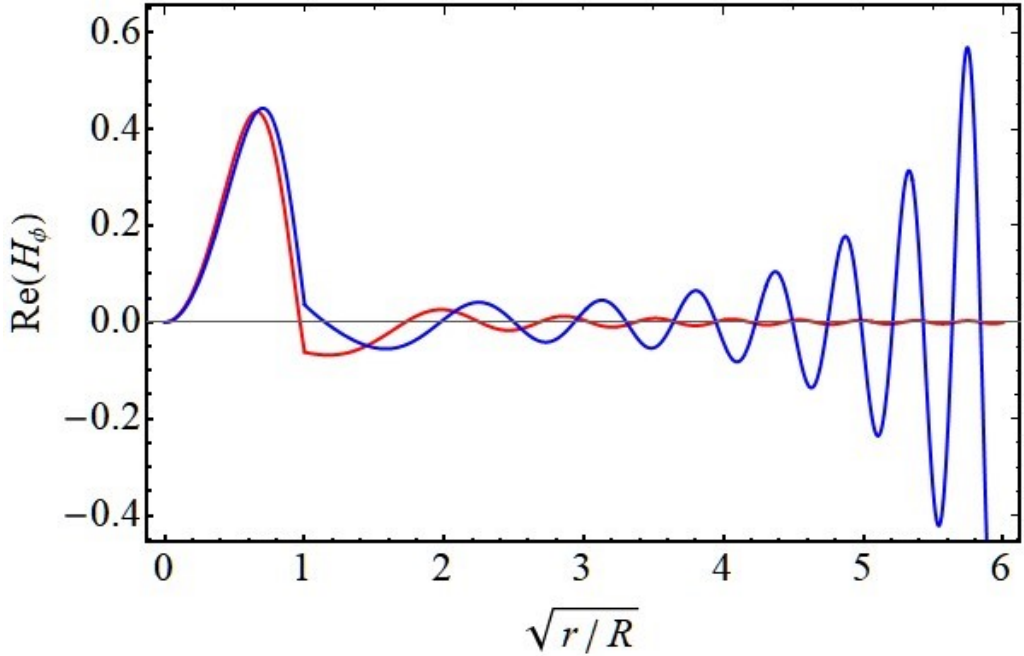

Fig. 1. The dependence of the magnetic field $\text{Re}(H_\varphi(r, \theta = \pi/2))$ on the radius for the usual TM$_{101}$ mode (blue curve, $k_0R=1.35715 - 0.160978i$) and for the perfect nonradiating PTM$_{101}$ mode ((7), red curve $k_0R=1.51893$) in a sphere with $\varepsilon = 10$.

Fig. 1 shows that the normal mode grows exponentially at infinity, while the perfect nonradiating mode goes to zero and has no radiation losses! Inside the resonator, the structures of these modes are similar.

Nonradiating modes are not the feature of spherical geometry and, apparently, exist for axisymmetric open resonators of an arbitrary shape. We have shown rigorously that such modes exist for arbitrary spheroids with semiaxes $a$ and $b$, having a volume equal to the volume of a sphere of the radius $R$ with a surface described by the equation

$$(\rho/b)^2 + (z/a)^2 = 1; a = Rt^{2/3}; b = Rt^{-1/3} \qquad (9),$$

where $t=a/b$. For $t < 1$, we have an oblate spheroid, and for $t > 1$, it is elongated.

The eigenfunctions and eigenfunctions of perfect nonradiating modes of such spheroids can be found by expanding the solutions of the equations (4) and (5) over spheroidal functions [38,39]. In the case of TM polarization for a prolate spheroid, the general axisymmetric solution of Maxwell's equations (4) and (5) in the prolate coordinate system $(1 < \xi < \infty, -1 < \eta < 1, 0 < \varphi < 2\pi)$ has the form:

$$H_{1,\varphi} = \sum_{n=1}^{\infty} a_n PS_{n1}(c_1,\eta) S_{n1}(c_1,\xi), \xi < \xi_0$$
$$H_{2,\varphi} = \sum_{n=1}^{\infty} b_n PS_{n1}(c_0,\eta) S_{n1}(c_0,\xi), 1 < \xi < \infty \qquad (10)$$

where $PS_{n1}(c,\eta)$ are the angular spheroidal functions, $S_{n1}(c,\xi)$ are the radial spheroidal functions of the first kind, $\xi_0 = a/\sqrt{a^2 - b^2}$ and $c_{0,1} = z_{0,1}\sqrt{t^2 - 1}/t^{1/3}$.

Equating the tangential components of the electric and magnetic fields on the surface of the spheroid (9) and using the orthogonality properties of angular spheroidal functions, one can find the dispersion equation describing the perfect nonradiating modes:

$$\det M = 0,$$
$$M_{np} = \Pi_{np}(c_1,c_0)\left(\varepsilon SD_p(c_0,\xi_0) S_{n1}(c_1,\xi_0) - S_{p1}(c_0,\xi_0) SD_n(c_1,\xi_0)\right) \qquad (11)$$

where

$$\Pi_{n,p}(c_1,c_0) = \int_{-1}^{1} d\eta\, PS_{n1}(c_1,\eta) PS_{p1}(c_0,\eta)$$

$$SD_n(c,\xi_0) = \frac{\partial\left(\left(\xi_0^2 - 1\right)^{1/2} S_{p1}(c,\xi_0)\right)}{\partial \xi_0}$$

Dispersion equation (11) is also valid for an oblate spheroid with the corresponding analytic continuation of spheroidal functions.

The exact solution of dispersion equations (11) for perfect TM modes (PTM) is shown in Fig. 2. This figure also shows the dispersion laws of usual TM modes and so-called confined modes [40] with magnetic field different from zero only inside the nanoparticle and complying with the equations:

$$\nabla \times \nabla \times \mathbf{H}_n = k_n^2 \mathbf{H}_n \quad \text{inside nanoparticle}$$
$$\mathbf{H}_n = 0 \quad \text{at the boundary} \tag{12}$$

and the electric field of the confined modes is zero everywhere.

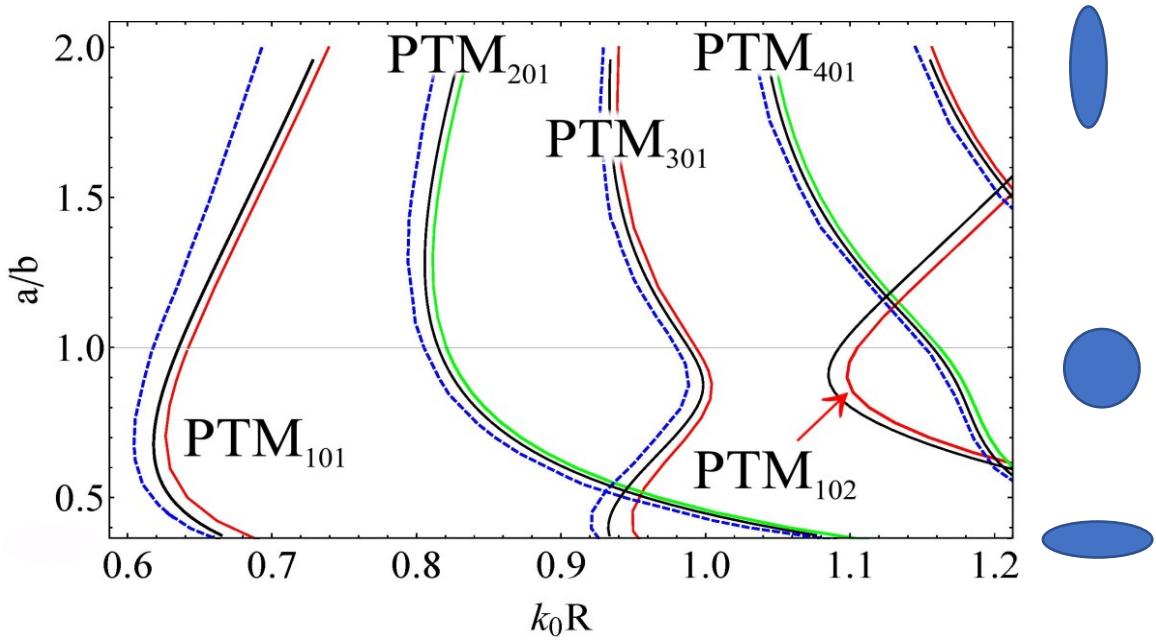

Fig.2. Perfect nonradiating TM modes (PTM) of spheroids with $\varepsilon=50$ as a function of the size parameter $k_0R$ and the spheroid aspect ratio, $a/b$. Red and green curves stand for odd and even perfect nonradiating modes, blue dashed lines correspond to usual normal modes. Black lines show frequencies of confined modes (12), $\omega_{n,confined} = k_n c / \sqrt{\varepsilon}$.

It can be seen from Fig. 2 that for each usual quasi-normal mode (blue dashed lines), there are its counterparts in the form of a perfect nonradiating mode (red and green lines) indicating that there are infinitely many perfect nonradiating modes. It should be emphasized here once again that the frequencies of the perfect nonradiating modes, the solution of (11), are real numbers!

Another interesting fact is that the eigenfrequencies of perfect nonradiating modes are substantially higher than the real parts of frequencies of usual modes $\omega_{n,usual}$. Moreover, it can be argued that the frequencies of confined modes (12), $\omega_{n,confined} = k_n c/\sqrt{\varepsilon}$ (black curves), appearing in the limit of infinite permittivity, are always situated between the frequencies of usual and perfect nonradiating modes, $\omega_{n,usual} < \omega_{n,confined} < \omega_{n,perfect}$. This relationship between frequencies is a manifestation of a very deep connection between confined modes (with fields localized strictly inside the resonator) and perfect nonradiating modes (with fields not localized inside the resonator). Fig. 3 shows the dependence of the frequencies $k_0 R\sqrt{\varepsilon}$ of usual and perfect nonradiating modes on the inverse permittivity.

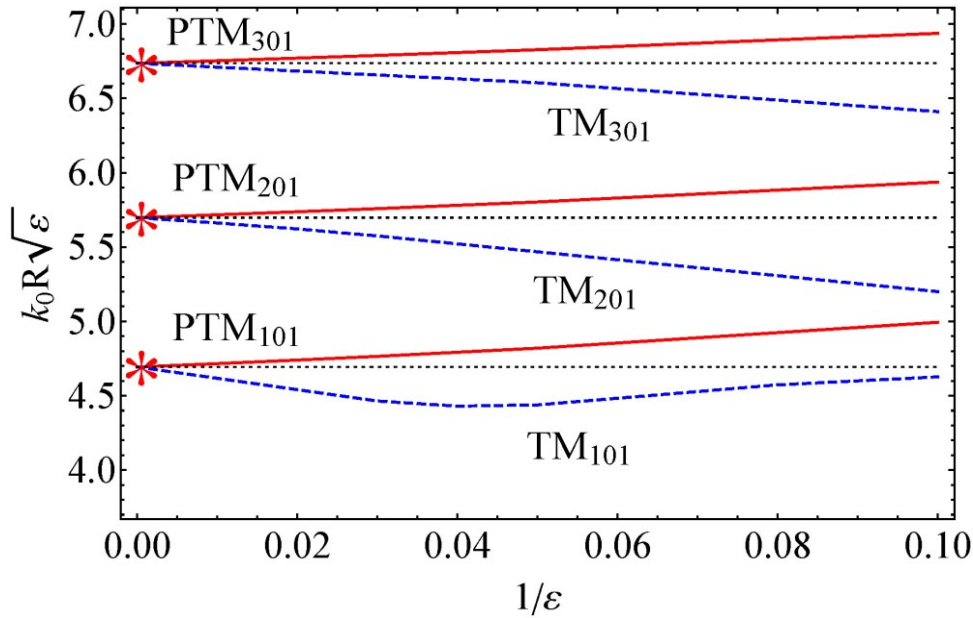

Fig. 3. Dependence of $k_0 R\sqrt{\varepsilon}$ for usual (blue dashed curves, TM) and perfect nonradiating modes (red curves, PTM) on $1/\varepsilon$ for a prolate spheroid with $a/b = 1.3$. The red asterisks and black dashed lines indicate confined modes (12).

From Fig. 3. it is seen that in the limit $\varepsilon \to \infty$ the usual and perfect nonradiating modes merge into confined ones. However, at finite permittivities, this degeneracy is lifted, and confined modes are split into perfect nonradiating modes with infinite $Q$-factors and the usual quasi-normal modes with finite $Q$-factors. From Fig.3 one can clearly see that there is nothing common between the frequencies of perfect nonradiating modes and the frequencies of usual normal modes.

Such an unambiguous connection between the frequencies of confined modes and perfect nonradiating modes allows us to assert that perfect nonradiating modes with TM polarization definitely exist for any axisymmetric dielectric bodies!

The found perfect nonradiating modes are not abstract solutions of sourceless Maxwell's equations. They are of great practical importance for finding the conditions for extremely small or even zero scattered power at a finite stored energy, leading to the unlimited radiative $Q$-factor. To demonstrate great practical importance of the perfect nonradiating modes, we have simulated within Comsol Multiphysics software the scattering of an axially symmetric Bessel beam by nanospheroids of different shapes. To observe the odd perfect nonradiating modes, we have used an odd incident Bessel beam:

$$H_\varphi \sim J_1(k_0 \sin\alpha\rho)\cos(k_0 z \cos\alpha) \qquad (13),$$

while to see the even perfect nonradiating modes, one should use an even incident Bessel beam:

$$H_\varphi \sim J_1(k_0 \sin\alpha\rho)\sin(k_0 z \cos\alpha) \qquad (14)$$

In (13), (14) $\rho$ and z are cylindrical coordinates and $\alpha$ is a conical angle of the Bessel beam. Let us stress that the Comsol simulations are fully independent on the analytical results (10), (11), and fully confirm our analytical results.

In Fig. 4 one can see the simulated dependence of scattered power $P_{rad}$, stored energy $W_{stored}$, and the generalized radiative quality factor $Q = \omega W_{stored}/P_{rad}$ on size parameter of nanospheroids for aspect ratio $a/b = 0.7$.

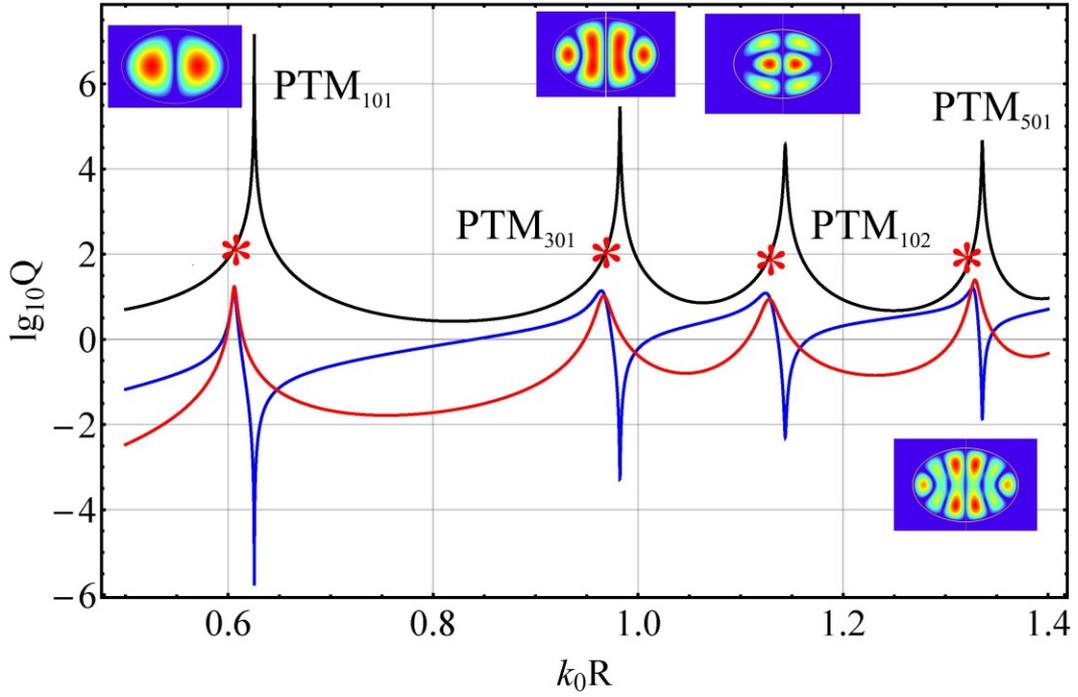

Fig.4. The dependence of scattered power $P_{rad}$ (blue), stored energy $W_{stored}$ (red), and generalized radiative $Q$ factor (black) on the size parameter of an oblate spheroid with $a/b = 0.7$ obtained within the Comsol simulation. TM symmetric excitation (13), $\alpha=\pi/4$, $\varepsilon=50$. The insets show the distribution of $|H_\varphi|$ in perfect nonradiating modes. All maxima correspond to perfect nonradiating modes! The red asterisks on the black curve show the $Q$-factor of the usual modes.

Figure 4 shows clearly the presence of perfect nonradiating modes, having $Q$-factors significantly higher than usual ones. It can be seen from this figure that, upon excitation (13), all the maxima of the generalized $Q$-factor are due to perfect nonradiating modes shown in Fig.2 and Fig.3. In this case, the $Q$-factors of usual modes (shown by asterisks) are several orders of magnitude lower than the $Q$-factors of perfect nonradiating modes! Moreover, smart optimization of the excitation beam [26] makes it possible to increase the $Q$ factor of perfect nonradiating modes almost unlimitedly.

Above, as an example, perfect nonradiating PTM modes in dielectric spheroidal nanoparticles are considered. However, this approach is directly generalized to TE polarization and to other axisymmetric nanoparticles [41]. Moreover, this approach is directly generalized to nanoparticles made from DNG or chiral metamaterials. However, the existence of perfect nonradiating modes for non-symmetric nanoparticles is still questionable.

In conclusion, we have put forward the concept of perfect nonradiating eigenmodes of light in dielectric nanoparticles. These modes are exact solutions of sourceless Maxwell equations. We have shown rigorously that in the case of axisymmetric particles such modes always exist and arise at frequencies of somewhat higher than resonance frequencies of usual modes.

Perfect nonradiating modes possess fully different physics than their weakly radiating counterparts and have no analogues. In particular, they differ from the so-called anapole modes [17-22,42] in that their field outside the particle is different from zero and has a well-defined expansion over spherical (2) or spheroidal (10) harmonics. These modes also differ from "bound states in a continuum" because they do not decay exponentially. The perfect modes are closest to strange Neumann-Wigner modes [43], but unlike the latter, the optical potential of a nanoparticle (permittivity) differs from vacuum value only in a finite region of space, fundamentally distinguishing perfect modes from Neumann-Wigner modes [43].

Due to extremely small scattered power and unlimited radiative $Q$-factors our finding paves the way for development of new nano-optical devices with high concentration of field inside nanoparticles and extremely small radiative losses, including low threshold nanolasers, biosensors, parametric amplifiers, and nanophotonics quantum circuits.

**Acknowledgments.** The reported study was funded by RFBR, project number 20-12-50136.

Supplemental material for the article
Perfect Nonradiating Modes in Dielectric Nanoparticles
Vasily Klimov
P.N. Lebedev Physical Institute, Russian Academy of Sciences, 53 Leninsky Prospekt,
Moscow 119991, Russia
E-mail: klimov256@gmail.com

## 1. Perfect nonradiating modes in dielectric sphere

Nonsingular solutions of the Maxwell's equations in a spherical coordinate system ($r, \theta, \varphi$) with TM polarization have the well-known form [1]:

$$E_r^{in} = -n(n+1)Y_{nm}\frac{j_n(k_1 r)}{k_1 r}e^{-i\omega t}$$

$$E_\theta^{in} = -\frac{\partial Y_{nm}}{\partial \theta}\frac{1}{k_1 r}\left[k_1 r j_n(k_1 r)\right]' e^{-i\omega t}$$

$$E_\varphi^{in} = -\frac{1}{\sin\theta}\frac{\partial Y_{nm}}{\partial \varphi}\frac{1}{k_1 r}\left[k_1 r j_n(k_1 r)\right]' e^{-i\omega t}$$

$$H_r^{in} = 0$$  (1),

$$H_\theta^{in} = -\frac{k_1}{i\omega\mu_1}\frac{1}{\sin\theta}\frac{\partial Y_{nm}}{\partial \varphi}j_n(k_1 r)e^{-i\omega t}$$

$$H_\varphi^{in} = \frac{k_1}{i\omega\mu_1}\frac{\partial Y_{nm}}{\partial \theta}j_n(k_1 r)e^{-i\omega t}$$

where $k_1$ stands for the wavenumber of media.

In the axisymmetric nonmagnetic case $m = 0$, $\mu_1=1$, there is no dependence on $\varphi$, and only the azimuthal component of the magnetic field $H_\varphi$ is nonzero, so instead of (1) inside the sphere one can write

$$E_r^{in} = -An(n+1)P_n(\cos\theta)\frac{j_n(k_1 r)}{k_1 r}e^{-i\omega t}$$

$$E_\theta^{in} = -A\frac{\partial P_n(\cos\theta)}{\partial \theta}\frac{1}{k_1 r}\left[k_1 r j_n(k_1 r)\right]' e^{-i\omega t}$$  (2).

$$H_\varphi^{in} = A\frac{k_1}{i\omega}\frac{\partial P_n(\cos\theta)}{\partial \theta}j_n(k_1 r)e^{-i\omega t}$$

$$E_\varphi^{in} = 0, H_r^{in} = 0, H_\theta^{in} = 0$$

Here, for the spherical harmonics with $m = 0$ we have used their expression in terms of the Legendre functions, $Y_{n0}(\theta,\varphi) = P_n(\cos\theta)$, $k_1 = \sqrt{\varepsilon}\, k_0$ is the wave number in the sphere, $j_n$ are the spherical Bessel functions, and $A$ is the amplitude of the perfect nonradiating mode inside the sphere.

Outside the sphere (in vacuum, $\varepsilon = 1$), we are looking for a solution in exactly the same nonsingular form (and this is a novelty!) with the replacement $k_1 \to k_0$:

$$E_r^{out} = -Bn(n+1)P_n(\cos\theta)\frac{j_n(k_0 r)}{k_0 r}e^{-i\omega t}$$

$$E_\theta^{out} = -B\frac{\partial P_n(\cos\theta)}{\partial \theta}\frac{1}{k_0 r}\left[k_0 r j_n(k_0 r)\right]' e^{-i\omega t} \quad (3).$$

$$H_\varphi^{out} = B\frac{k_1}{i\omega}\frac{\partial P_n(\cos\theta)}{\partial \theta}j_n(k_0 r)e^{-i\omega t}$$

$$E_\varphi^{out} = 0, H_r^{out} = 0, H_\theta^{out} = 0$$

In (3), $B$ is the amplitude of a perfect nonradiative mode outside the sphere.

On the boundary of the sphere $r = R$, the tangential components of the fields must be continuous:

$$H_\varphi^{in}(r=R) = H_\varphi^{out}(r=R)$$

$$E_\theta^{in}(r=R) = E_\theta^{out}(r=R) \quad (4),$$

that is, for any $0 < \theta < \pi$, the following conditions must be met:

$$A\frac{k_1}{i\omega}\frac{\partial P_n(\cos\theta)}{\partial \theta}j_n(k_1 R)e^{-i\omega t} = B\frac{k_0}{i\omega}\frac{\partial P_n(\cos\theta)}{\partial \theta}j_n(k_0 R)e^{-i\omega t}$$

$$-A\frac{\partial P_n(\cos\theta)}{\partial \theta}\frac{1}{k_1 R}\left[k_1 R j_n(k_1 R)\right]' e^{-i\omega t} = -B\frac{\partial P_n(\cos\theta)}{\partial \theta}\frac{1}{k_0 R}\left[k_0 R j_n(k_0 R)\right]' e^{-i\omega t} \quad (5).$$

Using the orthogonality of the Legendre polynomials, the system of equations that determine the eigenfrequencies of perfect nonradiating modes can be written as:

$$Ak_1 j_n(k_1 R) = Bk_0 j_n(k_0 R)$$

$$A\frac{1}{k_1 R}\left[k_1 R j_n(k_1 R)\right]' = B\frac{1}{k_0 R}\left[k_0 R j_n(k_0 R)\right]' \quad (6).$$

The nontrivial solutions (6), i.e. the perfect nonradiating modes, exist if the determinant of the system (6) is equal to zero:

$$\varepsilon j_n\left(\sqrt{\varepsilon}X\right)\left[Xj_n(X)\right]' - j_n(X)\left[\sqrt{\varepsilon}Xj_n\left(\sqrt{\varepsilon}X\right)\right]' = 0 \quad (7),$$

where $X = k_0 R$ is a so-called size parameter.

For each $\varepsilon > 1$ and $n > 1$, Eq. (7) has an infinite number of real roots, indicating the existence of perfect nonradiating modes, where the field distributions take the form:

$$H_\varphi^{in} = P_n^1(\cos\theta)j_n(k_1 r)e^{-i\omega t}, \text{inside sphere}, r < R$$

$$H_\varphi^{out} = P_n^1(\cos\theta)\frac{j_n(k_1 a)}{j_n(k_0 a)}j_n(k_0 r)e^{-i\omega t}, \text{outside sphere}, r > R \quad (8).$$

As an example, Fig. S1 shows the dependencies of the left side of (7) on the size parameter $X$ for a sphere with permittivity $\varepsilon = 10$.

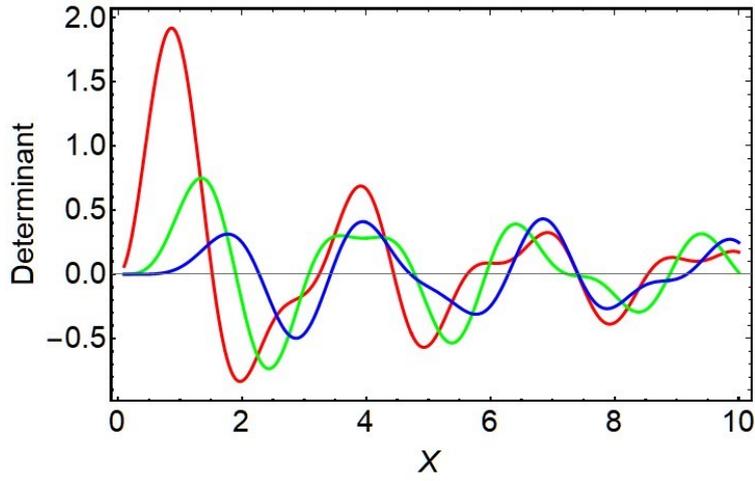

Fig. S1. Dependences of the left side of (7) on the size parameter $X=k_0R$ for a sphere with permittivity $\varepsilon = 10$ for $n = 1$ (red), $n = 2$ (green), $n = 3$ (blue).

Fig. S1 shows an infinite set of real roots that correspond to the perfect nonradiating modes. The dependence of the radial part of usual and perfect non-radiating modes on the radius is shown in Fig. 1 of the manuscript.

Fig. S2 shows the spatial distribution of $\mathrm{Re}H_\varphi$ in the perfect nonradiating mode and the usual mode in a sphere with $\varepsilon = 10$.

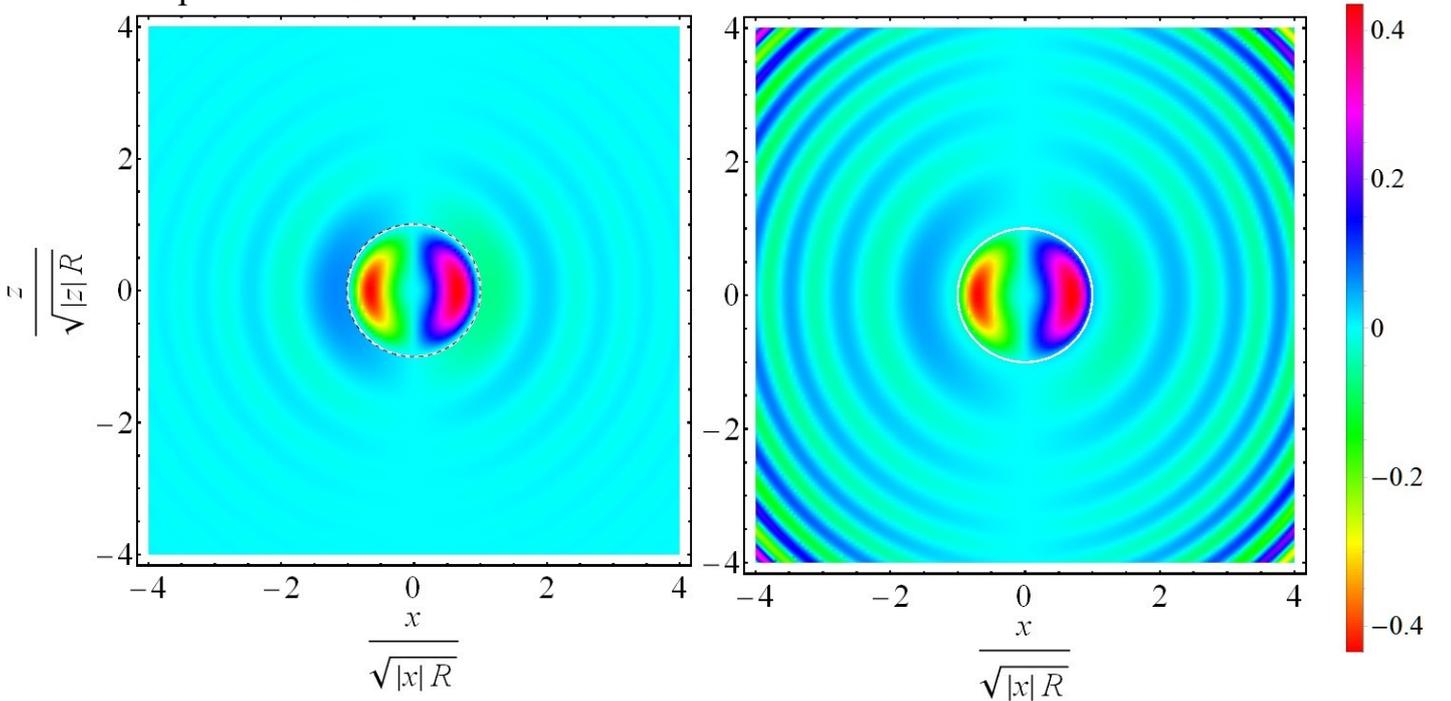

Fig. S2. Spatial distribution of $\mathrm{Re}H_\varphi$ in the perfect nonradiating $\mathrm{PTM}_{101}$ mode ($k_0R==1.51893$, left) and the usual $\mathrm{TM}_{101}$ mode ($k_0R=1.35715 - 0.160978\,i$, right) in a sphere of the radius $R$ with $\varepsilon = 10$.

It is clearly seen from Fig. S2 that the perfect mode exists, has a sensible spatial distribution, and decreases at infinity, while the usual mode increases exponentially at infinity.

Thus, the existence of the perfect nonradiating modes in the dielectric sphere has been rigorously proved. To observe TM perfect nonradiating modes in the sphere, we have performed the Comsol simulations (see Fig.S3) with an incident Bessel beam:

$$H_\varphi \sim J_1(k_0 \sin\alpha \rho) e^{ik_0 z \cos\alpha} \qquad (9)$$

In (9) $\rho$ and $z$ are cylindrical coordinates and $\alpha$ is a conical angle of the Bessel beam.

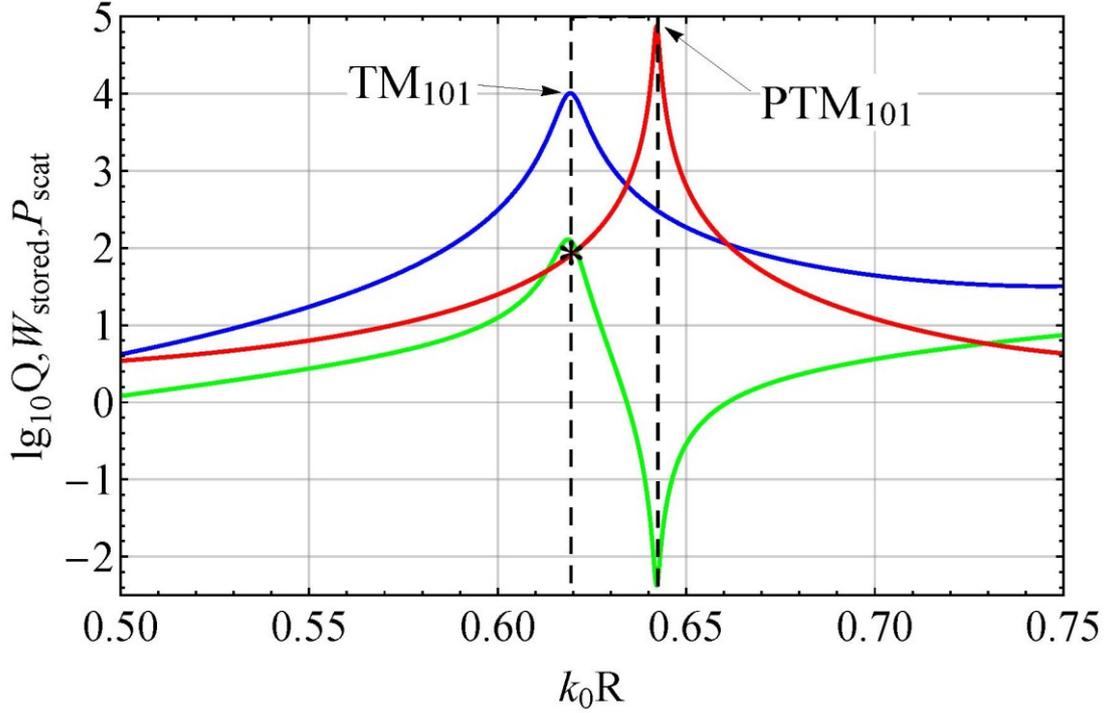

Fig.S3. The dependence of scattered power $P_{rad}$ (green), stored energy $W_{stored}$ (blue), and generalized radiative quality factor $Q = \omega W_{stored} / P_{rad}$ (red) on the size parameter of a sphere obtained within the Comsol simulation. TM excitation (9), $\alpha=\pi/4$, $\varepsilon=50$. The asterisk on the red curve shows the $Q$ factor value of the $TM_{101}$ mode.

Fig.S3 shows clearly the appearance of an extremely high quality perfect $PTM_{101}$ mode in the spectra of a Bessel beam (9) scattered by a sphere. Note that $Q$ factor of the $PTM_{101}$ mode is by three orders of magnitude greater than $Q$ factor of a usual $TM_{101}$ mode!

2. Perfect nonradiating modes in dielectric spheroids

Apparently, perfect nonradiating modes exist for axisymmetric bodies of an arbitrary shape. It is shown strictly below that such modes exist for arbitrary spheroids with semiaxes $a$ and $b$, having a volume equal to the volume of a sphere of the radius $R$ and a surface described by the equation:

$$\left(\rho/b\right)^2 + \left(z/a\right)^2 = 1; a = Rt^{2/3}; b = Rt^{-1/3}, \quad (10),$$

where $t=a/b$. For $t < 1$, we have an oblate spheroid, and for $t > 1$, it is elongated.

The eigenfunctions and eigenfunctions of perfect nonradiating modes of such spheroids can be found by solving sourceless Maxwell's equations in the prolate spheroidal coordinates $\xi, \eta, \varphi$ [2,3], related to Cartesian coordinates by the equations:

$$x = \frac{d}{2}\left(\xi^2 - 1\right)^{1/2}\left(1 - \eta^2\right)^{1/2}\cos\varphi, y = \frac{d}{2}\left(\xi^2 - 1\right)^{1/2}\left(1 - \eta^2\right)^{1/2}\sin\varphi, z = \frac{d}{2}\xi\eta; \quad (11)$$

where $d = 2\sqrt{a^2 - b^2}$.

In these coordinates, the surface of the spheroid (10) is determined by the condition:

$$\xi = \xi_0 = a/\sqrt{a^2 - b^2}$$

In prolate spheroidal coordinates, the Lamé coefficients have the form [3]:

$$h_\xi = \frac{d}{2}\sqrt{\frac{\xi^2 - \eta^2}{\xi^2 - 1}}; h_\eta = \frac{d}{2}\sqrt{\frac{\xi^2 - \eta^2}{1 - \eta^2}}; h_\varphi = \frac{d}{2}\sqrt{\left(\xi^2 - 1\right)\left(1 - \eta^2\right)} \quad (12)$$

In spheroidal coordinates, the variables can be separated, and solutions of Maxwell's equations can be represented as an expansion over spheroidal wave functions.

2.1 TM polarization, non-magnetic case

In the case of TM polarization, for a single nonzero component of the magnetic field, we can write

$$H_{1,\varphi} = \sum_{n=1}^{\infty} a_n PS_{n1}(c_1, \eta) S_{n1}(c_1, \xi), c_1 = k_0 R\sqrt{\varepsilon}\sqrt{t^2 - 1}/t^{1/3} \text{ inside nanoparticle}, 1 < \xi < \xi_0$$

$$H_{2,\varphi} = \sum_{n=1}^{\infty} b_n PS_{n1}(c_0, \eta) S_{n1}(c_0, \xi), c_0 = k_0 R\sqrt{t^2 - 1}/t^{1/3} \text{ everywhere}, \xi > 1 \quad (13)$$

where $PS_{n1}(c, \eta)$ are the angular spheroidal functions, $S_{n1}(c, \xi)$ are the radial spheroidal functions of the first kind [2].

The tangential component of the electric field

$$E_\eta = \frac{i}{k_0\varepsilon}\frac{1}{h_\xi h_\varphi}\frac{\partial h_\varphi H_\varphi}{\partial \xi} \quad (14)$$

looks like:

$$E_{1,\eta} = \frac{1}{\varepsilon} \frac{2i}{k_0 d \left(\xi^2 - \eta^2\right)^{1/2}} \sum_{n=1}^{\infty} a_n PS_{n1}(c_1, \eta) \frac{\partial \left(\xi^2 - 1\right)^{1/2} S_{n1}(c_1, \xi)}{\partial \xi} \tag{15}$$

inside nanoparticles, $\xi < \xi_0$, and

$$E_{2,\eta} = \frac{2i}{k_0 d \left(\xi^2 - \eta^2\right)^{1/2}} \sum_{n=1}^{\infty} b_n PS_{n1}(c_0, \eta) \frac{\partial \left(\xi^2 - 1\right)^{1/2} S_{n1}(c_0, \xi)}{\partial \xi} \tag{16}$$

everywhere, $\xi > 1$.

After multiplication by angular harmonics, integration over $\eta$, and application of the orthogonality condition for angular spheroidal functions,

$$\int_{-1}^{1} d\eta\, PS_{n1}(c_1, \eta) PS_{m1}(c_1, \eta) = \delta_{nm} NN_n = \delta_{nm} \frac{2n(n+1)}{2n+1} \tag{17}$$

the conditions for the continuity of the magnetic field at the spheroid boundary $\xi = \xi_0$ take the form:

$$a_n NN_n S_{n1}(c_1, \xi_0) = \sum_{p=1}^{\infty} \Pi_{np}(c_1, c_0) S_{p1}(c_0, \xi_0) b_p \tag{18}$$

Similarly, the boundary conditions for the continuity of the electric field at the boundary of the spheroid $\xi = \xi_0$ can be written in the form:

$$\frac{1}{\varepsilon} a_n NN_n SD_n(c_1, \xi_0) = \sum_{p=1}^{\infty} \Pi_{np}(c_1, c_0) SD_p(c_0, \xi_0) b_p, \tag{19}$$

where

$$SD_n(c, \xi_0) = \frac{\partial \left(\left(\xi_0^2 - 1\right)^{1/2} S_{p1}(c, \xi_0)\right)}{\partial \xi_0} \tag{20}$$

In (18) and (19)

$$\Pi_{n,p}(c_1, c_0) = \int_{-1}^{1} d\eta\, PS_{n1}(c_1, \eta) PS_{p1}(c_0, \eta) \tag{21}$$

stands for the overlap integral of angular spheroidal functions for different $c_1$ and $c_0$.

Fig.S4 shows the dependence of the overlap integral of angular spheroidal functions $\Pi_{n,p}(c_1, c_0)$ on the indices $n, p$.

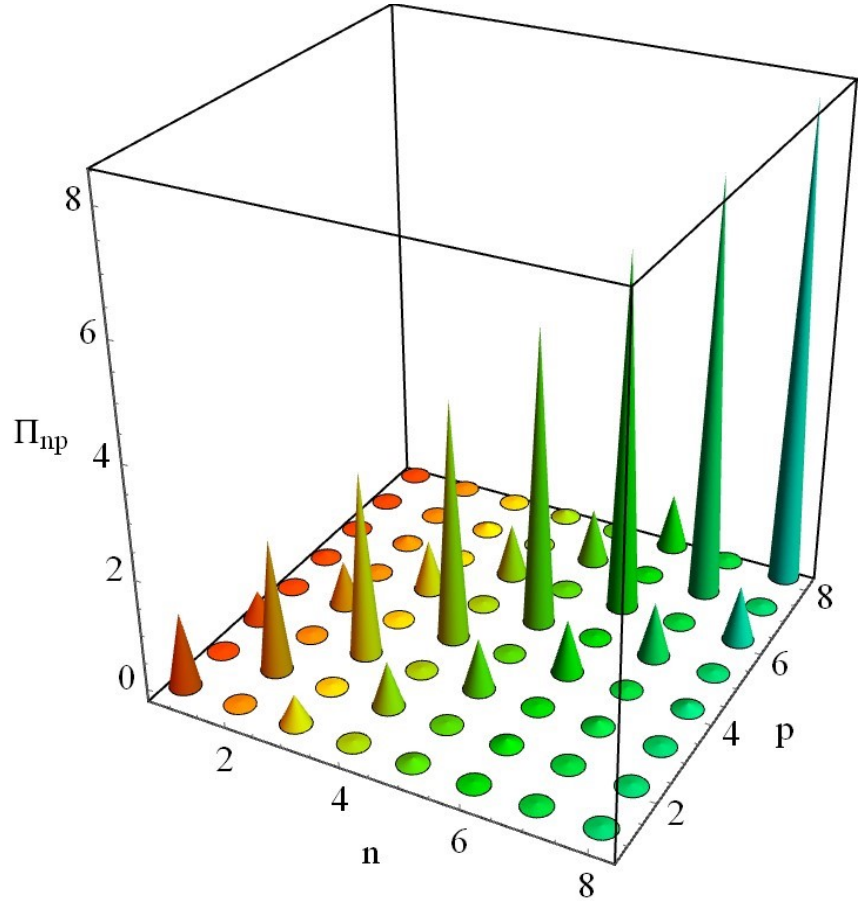

Fig.S4. Dependence of the overlap integral of angular spheroidal functions in an elongated spheroid $\Pi_{n,p}(c_1,c_0)$ on indices. $c_1 = 4, c_0 = c_1/\sqrt{\varepsilon} = 0.5657, \varepsilon = 50$

Fig.S4 shows that:

1) only modes with the same parity interact with each other;

2) for each mode, interaction is essential only with the nearest modes of the same parity, $2k \Leftrightarrow 2(k \pm 1); 2k+1 \Leftrightarrow 2(k \pm 1)+1$. This circumstance simplifies calculations since matrices of finite dimension 3×3 can be used to calculate eigenfrequencies.

Eliminating $a_n$ from (18) and (19), we obtain a homogeneous system for the coefficients $b_p$, determining the magnetic field outside the particle:

$$\sum_{p=1}^{\infty} \Pi_{np}(c_1,c_0)\left(\varepsilon SD_{p1}(c_0,\xi_0)S_{n1}(c_1,\xi_0) - S_{p1}(c_0,\xi_0)SD_{n1}(c_1,\xi_0)\right)b_p = 0 \qquad (22),$$

The compatibility condition of (22) allows one to find the modes and eigenfrequencies of the perfect nonradiating modes shown in Fig. 2 of the article.

For example, the solution of (13),(18) and (22) for PTM$_{101}$ mode in a prolate spheroid with $\varepsilon=10$ and a/b=1.3 is as follows:

$k_0R=1.57757$,

$a_1= -7.22705, a_3 = 0.0924, a5=0.004323, a_7= 0.00019$

$b_1 = 1, b_3 = -0.59122, b_5 = 1.08884, b_7 = -2.81192$

and is shown in Fig.S5.

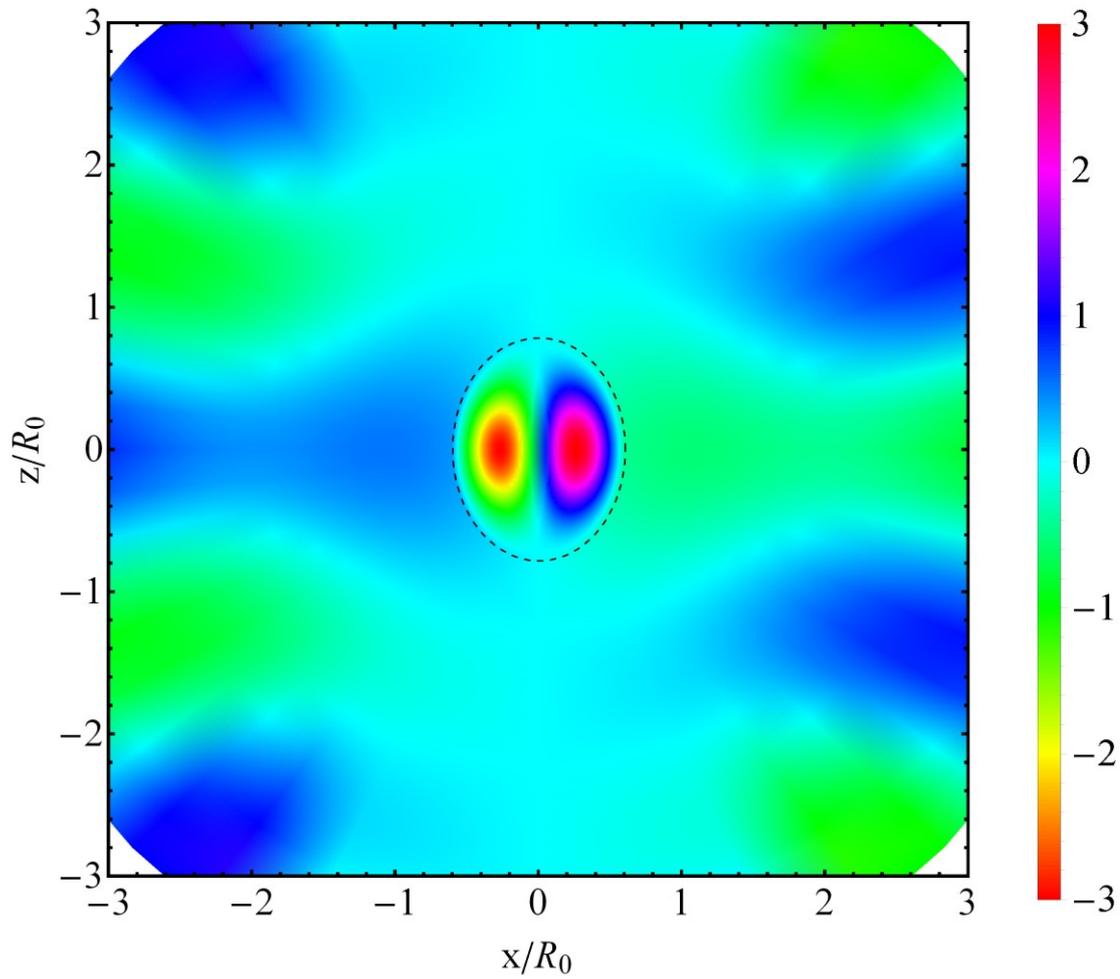

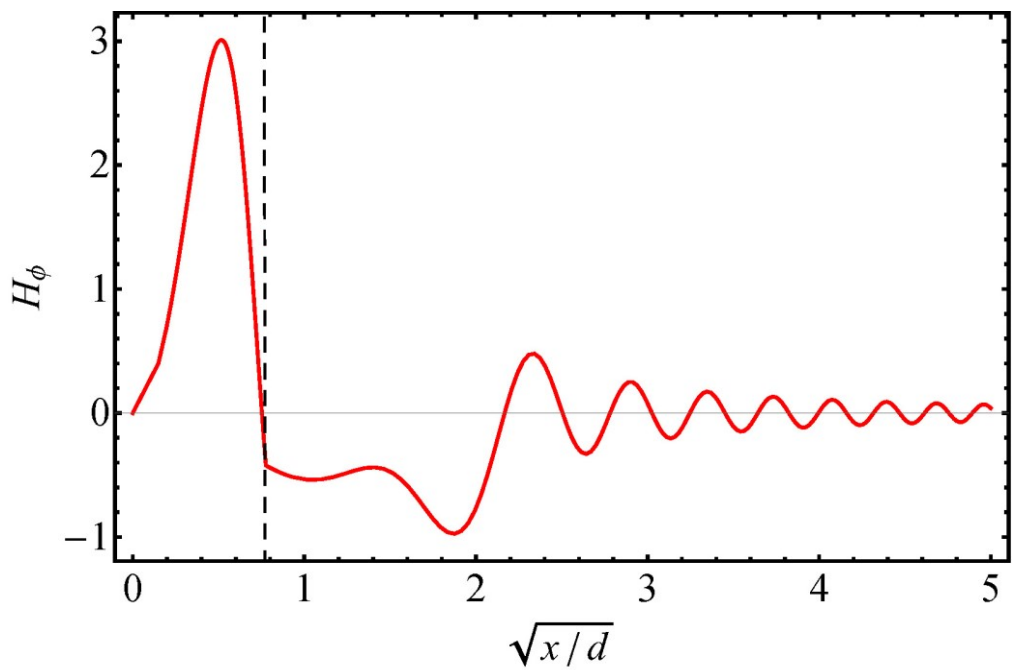

Fig. S5. Spatial distribution of $H_\varphi(x,z)$ (a) and the dependence of $H_\varphi(x,z=0)$ on $x$ (b) in the PTM$_{101}$ mode in a prolate spheroid with $\varepsilon=10$, $k_0R=1.57757$, $k_0d=2.4$, $a/b=1.3$.

Comparison of Fig.S5 and Fig.S2 (left) and Fig.1 shows that the structure of a perfect nonradiating mode in a spheroid looks generally similar to the structure of a perfect nonradiating mode in a sphere.

2.2. TE polarization, non-magnetic case.

In this case, the only nonzero component of the electric field can be written in the form:

$$E_{1,\varphi} = \sum_{n=1}^{\infty} a_n PS_{n1}(c_1,\eta) S_{n1}(c_1,\xi), c_1 = k_0 R\sqrt{\varepsilon}\sqrt{t^2-1}/t^{1/3} \text{ inside nanoparticle}, \xi<\xi_0$$

$$E_{2,\varphi} = \sum_{n=1}^{\infty} b_n PS_{n1}(c_0,\eta) S_{n1}(c_0,\xi), c_0 = k_0 R\sqrt{t^2-1}/t^{1/3} \quad \text{everywhere}, \xi>1 \tag{23}$$

Repeating the reasoning for TM polarization, for the coefficients $b_p$, determining the electric field outside the particle, we obtain a homogeneous system of equations:

$$\sum_{p=1}^{\infty} \Pi_{np}(c_1,c_0)\left(SD_{p1}(c_0,\xi_0)S_{n1}(c_1,\xi) - S_{p1}(c_0,\xi_0)SD_{n1}(c_1,\xi_0)\right)b_p = 0 \tag{24}$$

that differs from the dispersion equation (22) only by the absence of $\varepsilon$ in the parenthesis. The compatibility condition of (24) allows one to find the eigenfrequencies of perfect nonradiating modes with TE polarization (see Fig. S6).

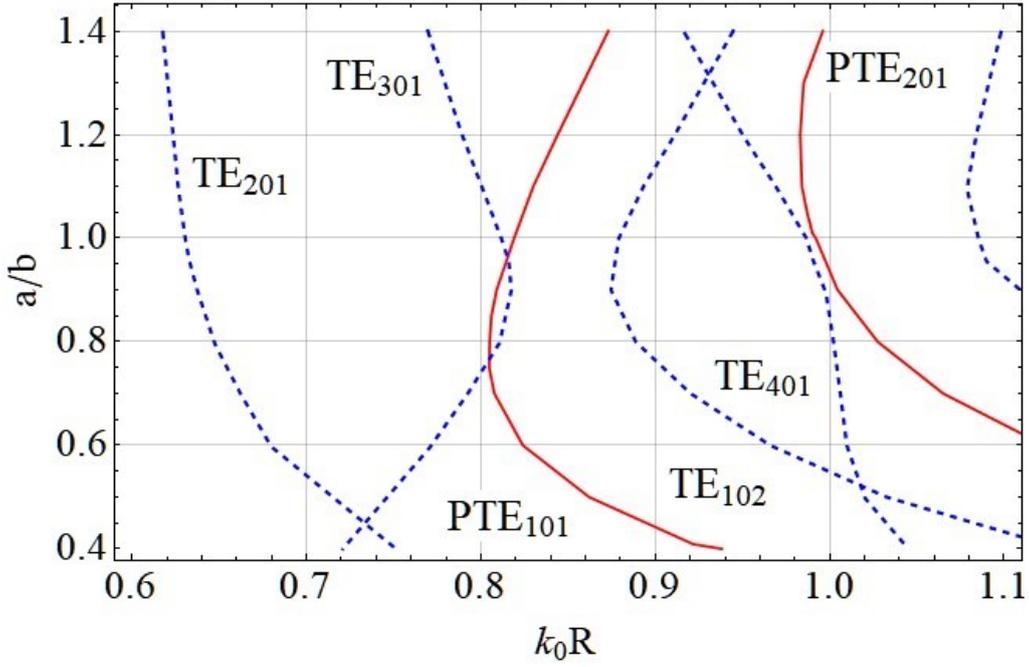

Fig.S6. Perfect nonradiating TE modes of spheroids with $\varepsilon=50$ as a function of the size parameter $k_0R$ and the aspect ratio, $a/b$. Red curves stand for TE perfect nonradiating modes (PTE), blue dashed lines correspond to usual TE modes.

From Fig.S6 and the structure of dispersion equation (24), it follows that the number of perfect nonradiating modes with TE polarization is infinite, as in the case of TM polarization.

It is clearly seen from Fig. S6, that the frequencies of the perfect nonradiating modes with TE polarization are located very far from the frequencies of the usual modes of the similar symmetry, indicating that there is no connection between usual and perfect modes.

Demonstration of the existence of TE perfect nonradiating modes in scattering spectra is more complicated in comparison with the TM case, since perfect modes with TE polarization are not confined modes [4] in the limit $\varepsilon \to \infty$. In addition, the frequencies of perfect $PTE_{n,0,1}$ modes turn out to be close to the frequencies of usual $TE_{n+1,0,1}$ modes. Therefore, to demonstrate perfect nonradiating TE modes, an incident beam should not excite usual nearby modes. In particular, to demonstrate the existence of the $PTE_{101}$ mode (see Fig.S7), it is necessary to suppress the excitation of the usual $TE_{301}$ mode, by using an exciting field of the form:

$$E_\varphi(r,\theta) \sim \left(j_1(k_0r)P_1^1(\theta) + \xi\, j_3(k_0r)P_3^1(\theta)\right) \qquad (25)$$

In (25), $r$ and $\theta$ are spherical coordinates, $j_n(z)$ and $P_n^1(\cos\theta)$ are the spherical Bessel functions and the Legendre polynomial, correspondingly.

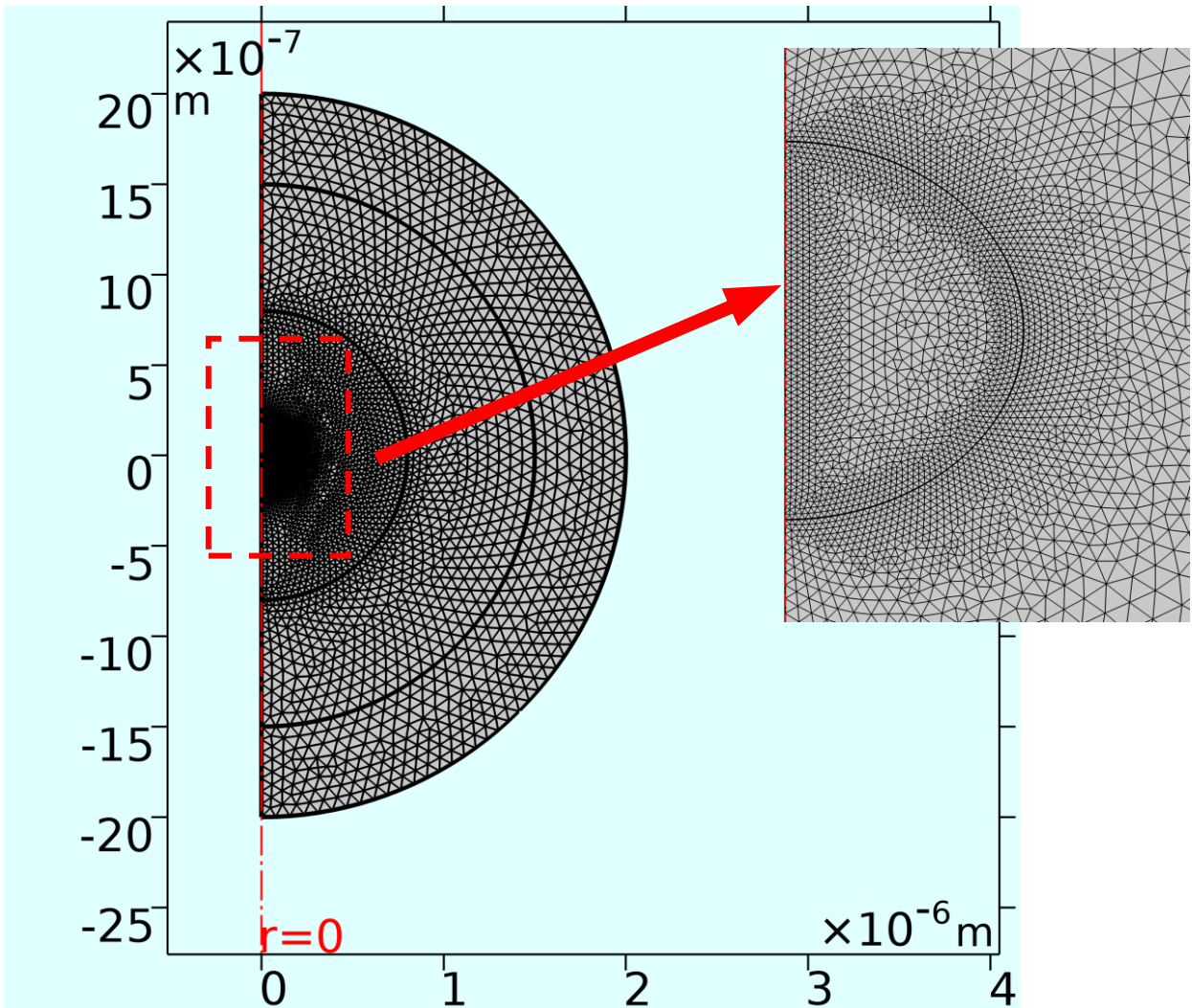

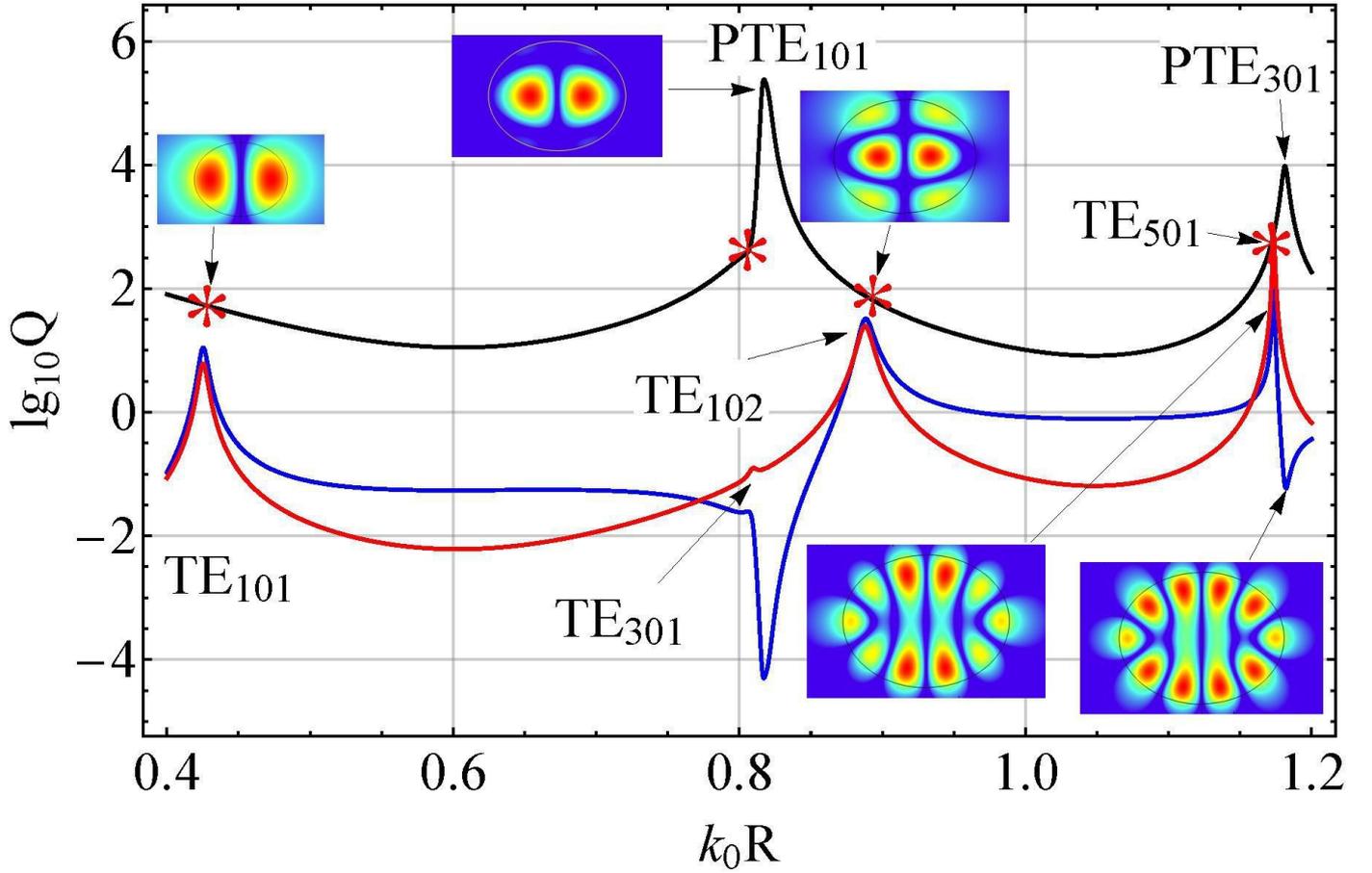

Fig.S7. a) 2D Comsol mesh used in simulations; b) The dependence of scattered power (blue), stored energy (red), and generalized Q-factor (black) on the size parameter of nanoparticles simulated within Comsol Multiphysics. TE excitation (25), $\xi$=-9.915, $\varepsilon$=50, $a/b$=0.8 (oblate spheroid). The insets show the distribution of $|H_\varphi|$ corresponding to the maxima of stored energy or Q-factor. The asterisks on the black curve show the Q factor values of the usual modes.

Fig.S7 clearly shows the presence of the perfect nonradiating modes, having Q-factors significantly higher than the Q-factors of usual modes. Moreover, smart optimization of the excitation beam [5] makes it possible to increase the Q-factor of the perfect nonradiating modes unlimitedly (with neglect of Joule losses in the nanoparticle material, of course).

### 2. Perfect nonradiating modes in a cylinder: TM polarization

We have not yet succeeded in finding an analytical solution for perfect nonradiating modes in a dielectric cylinder of a finite height. However, proceeding from the very plausible hypothesis that the existence of perfect modes is associated with the existence of confined modes, we have found the frequencies of perfect nonradiating modes by the Comsol simulation of the scattering of a Bessel beam

$$H_\varphi \sim J_1(k_0 \sin\alpha\rho)\cos(k_0 z \cos\alpha) \quad (26)$$

by a superspheroid with the surface described by the equation:

$$(\rho/a(t))^t + (z/c(t))^t = 1 \quad (27)$$

For $t = 2$, this is a sphere: $a(2) = c(2) = R$:

$$\rho^2/R^2 + z^2/R^2 = 1 \quad (28)$$

At $t = \infty$, this is a cylinder with the diameter $2a(\infty) = D_{cyl}$ and the height $H_{cyl} = 2c(\infty)$. The results of the Comsol simulation are shown in Fig. S8.

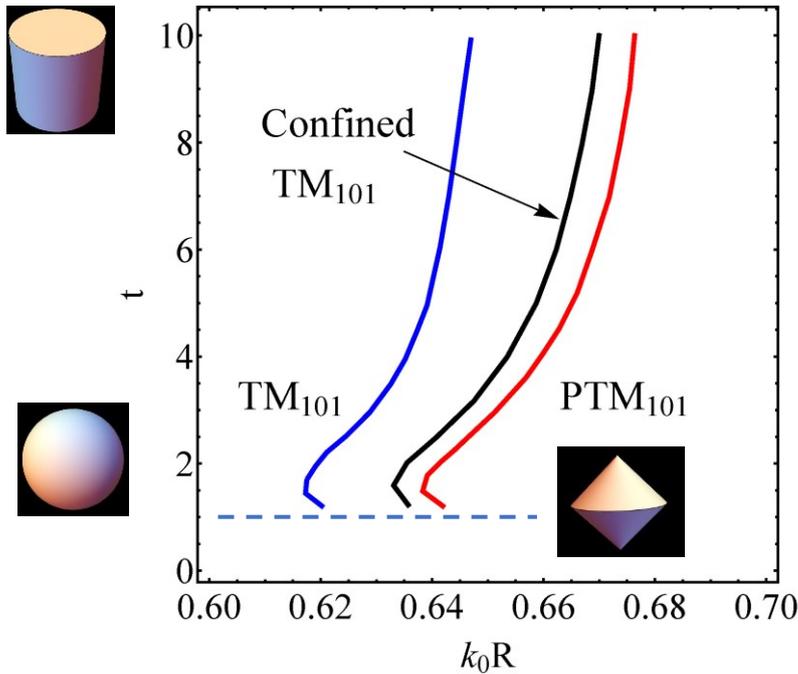

Fig.S8. Perfect nonradiating PTM modes of superspheroids with $\varepsilon=50$ as a function of the size parameter, $k_0R$ and the shape parameter, $t$. Red curve stands for $PTM_{101}$ perfect nonradiating mode (PTM), blue line corresponds to usual $TM_{101}$ mode. Black line corresponds to confined ($\varepsilon\to\infty$) $TM_{101}$ mode [4].

It is clearly seen from Fig. S8, that the frequencies of the perfect nonradiating modes with TM polarization in superspheroids are located very far from the frequencies of the usual modes of the similar symmetry, indicating that there is no connection between usual and perfect modes.

In Fig. S9, one can see the dependence of scattered power, stored energy, and generalized $Q$-factor on the size parameter $k_0R$ of a superspheroid with shape parameter $t = 8$, when excited by a Bessel beam (26).

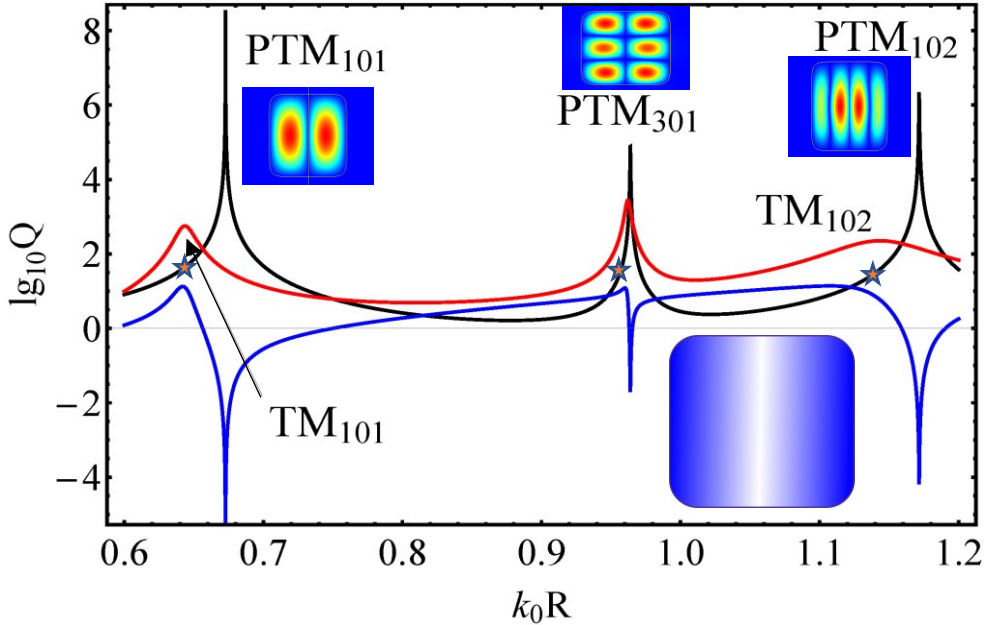

Fig.S9. The dependence of scattered power (blue), stored energy (red), and $Q$-factor (black) on the size parameter of a superspheroid with D/H=0.96, $t$=8 obtained as a result of the Comsol simulation .TM symmetric excitation (26), $\varepsilon$=50, $\alpha=\pi/4$ . The insets show the field distribution of the perfect nonradiating modes and the shape of the investigated superspheroid. All maxima correspond to perfect nonradiating modes! The asterisks on the black curve show the $Q$-factor values of the usual modes.

Fig. S9 shows that all the maxima of the generalized $Q$-factor with the excitation (26) are due to perfect nonradiating modes, with $Q$-factors of several orders of magnitude higher than the $Q$-factors of usual modes.